# STATISTICAL ANALYSIS OF STELLAR EVOLUTION[1]

By David A. van Dyk, Steven DeGennaro, Nathan Stein,
William H. Jefferys and Ted von Hippel

*University of California, Irvine, University of Texas at Austin, Harvard
University, University of Vermont and University of Texas at Austin,
and Siena College and University of Miami*

Color-Magnitude Diagrams (CMDs) are plots that compare the magnitudes (luminosities) of stars in different wavelengths of light (colors). High nonlinear correlations among the mass, color, and surface temperature of newly formed stars induce a long narrow curved point cloud in a CMD known as the main sequence. Aging stars form new CMD groups of red giants and white dwarfs. The physical processes that govern this evolution can be described with mathematical models and explored using complex computer models. These calculations are designed to predict the plotted magnitudes as a function of parameters of scientific interest, such as stellar age, mass, and metallicity. Here, we describe how we use the computer models as a component of a complex likelihood function in a Bayesian analysis that requires sophisticated computing, corrects for contamination of the data by field stars, accounts for complications caused by unresolved binary-star systems, and aims to compare competing physics-based computer models of stellar evolution.

**1. Introduction.** For most of their lives, stars are powered by thermonuclear fusion in their cores. In this process multiple atomic particles join together to form a heavier nucleus and energy is released as a byproduct. As this process continues for millions or billions of years, depending on the initial mass of the star, the composition of the star changes. When these

Received June 2008; revised October 2008.
[1]Supported in part by NSF Grants DMS-04-06085 and AST-06-07480 and by the Astrostatistics Program hosted by the Statistical and Applied Mathematical Sciences Institute. Any opinions, findings, and conclusions or recommendations expressed in this material are those of the author(s) and do not necessarily reflect the views of the National Science Foundation.

changes become severe enough to significantly affect the physical processes at the core, dramatic shifts in the color, spectrum, and density of the star occur that have long been observed by astronomers. In the early twentieth century two astronomers, Ejnar Hertzsprung and Henry Norris Russell, produced plots comparing the luminosity (energy radiated per unit time) and effective surface temperature of stars. Today generalizations of these plots are commonly called Color-Magnitude Diagrams (CMDs) and can be used to clearly separate out groups of stars powered by different physical processes and at different stages of their lives. These groups include the *main sequence*, so named for its dominant position in a CMD, the evolved red giants, and the even older white dwarfs. Today the physical processes that govern stellar formation and evolution are studied with complex computer models that can be used to predict the plotted magnitudes on a set of CMDs as a function of stellar parameters of interest, such as distance, stellar age, initial mass, and metallicity (a measure of the abundance of elements heaver than helium). *Luminosity* is a direct measurement of the amount of energy an astronomical object radiates per unit time, while a *magnitude* is a negative logarithmic transformation of luminosity; thus, smaller magnitudes correspond to brighter objects. In this paper we describe how we use these computer-based *stellar evolution models* as a component in a complex likelihood function and how we use Bayesian methods to fit the resulting statistical model. Thus, our aim is to fit physically meaningful stellar parameters and compare stellar evolution models by developing principled statistical methods that directly incorporate the evolution models via state-of-the-art complex computer models.

We focus on developing methods for the analysis of CMDs of the stars in a so-called *open cluster*. Stars in these clusters were all formed from the same molecular cloud at roughly the same time and reside as a physical cluster in space. This simplifies statistical analysis because we expect the stars to have nearly the same metallicity, age, and distance; only their masses differ. Unfortunately, the data are contaminated with stars that are in the same line of sight as the cluster but are not part of the cluster. These stars appear to be in the same field of view and are called *field stars*. Because field stars are generally of different ages, metallicities, and distances than the cluster stars, we are unable to constrain the values of these parameters and, thus, their coordinates on the CMDs are not well predicted from the computer models. The solution is to treat the data as a mixture of cluster stars and field stars, in which field stars are identified by their discordance with the model for the cluster stars. A second complication arises from multi-star systems in the cluster. These stars are the same age and have the same metallicity as the cluster, but we typically cannot resolve the individual stars in the system and thus observe only the sums of their luminosities in different colors. This causes these systems to appear systematically offset from the main sequence in a CMD. Because the offset is informative as to the individual stellar masses, however, we can formulate a statistical model to identify the individual masses.

Owing to the complexity of the computer-based stellar evolution models, the posterior distribution for the parameters of scientific interest under our statistical model is highly irregular. There are very strong and sometimes highly nonlinear correlations among the parameters. Some two-dimensional marginal distributions appear to be degenerate, with their probability mass lying completely on a one-dimensional curve. Sophisticated Markov chain Monte Carlo (MCMC) methods are required to explore these distributions. Our strategy involves dynamically transforming the parameters with the aim of reducing correlations. We use initial runs of the MCMC sampler to diagnose the correlations and automatically construct transformations that are used in a second run allowing the modified MCMC sampler to explore the posterior distribution. We are also developing methods to evaluate our statistical model and its underlying stellar evolution models with the ultimate goal of comparing and evaluating the physics-based computer models of stellar evolution.

Our use of principled statistical models and methods stands in contrast to the more ad-hoc methods that are often employed. A typical strategy for arriving at values for stellar parameters using the computer-based stellar evolution models involves over-plotting the data with the model evaluated at a set of parameter values and manually adjusting the values in order to visually improve the correspondence between the model and the data [e.g., Caputo et al. (1990); Montgomery, Marschall and Janes (1993); Dinescu et al. (1995); Chaboyer, Demarque and Sarajedini (1996); Rosvick and Vandenberg (1998); Sarajedini et al. (1999); VandenBerg and Stetson (2004)]. Experience leads to intuition as to which parameter should be adjusted in what way to correct for a particular discrepancy between the data and the model. Nonetheless, it is difficult to be sure one has found the optimal fit or to access the statistical error in the fit. To compare competing models, some researchers simulate data sets under each model with stellar parameters fit in this way. The simulated data sets are then compared with the actual data by comparing star counts in each bin of a grid superimposed on the CMD [e.g., Gallart et al. (1999); Cignoni et al. (2006)]. Other researchers have calculated the marginal distributions of stars on both axes of the CMD, comparing observed and simulated distributions in color and luminosity [e.g., Tosi et al. (1991, 2007)]. We are aware of one other group [Hernandez and Valls-Gabaud (2008)] applying an approach broadly similar to ours, though their technical approach and their scientific goals are meaningfully different than ours; see DeGennaro et al. (2008). Compared to the classical eyeball fitting of model to the data and compared to the statistical techniques developed to date, we believe that our principled statistical methods offer a more precise and reliable exploration of the parameters of stellar evolution.

The remainder of the paper is organized into five sections. We begin in Section 2 by outlining the relevant scientific background on stellar evolution models, their computational implementations, and the data available for fitting the models. Section 3 describes our formulation of a statistical model that incorporates the computer models while accounting for measurement error, binary-star systems, and field-star contamination. Statistical computation is discussed in Section 4, including our dynamic methods for improving efficiency. Analysis of the Hyades cluster is described in Section 5, followed by discussion in Section 6.

## 2. Stellar evolution.

2.1. *Basic evolutionary model and color-magnitude diagrams.* Stars are believed to be formed when the dense parts of a molecular cloud collapse into a ball of plasma. If the mass of the resulting *protostar* exceeds about 10% of the mass of the Sun, $M_\odot$, its core will ignite in a thermonuclear reaction that is powered by the fusion of hydrogen into helium. This reaction at the star's core can continue for millions or billions of years depending on the original mass and composition of the star. More massive stars are denser, and thus hotter, and burn their fuel more quickly. When the hydrogen at the core has been mostly converted into helium, the core collapses and the inner temperature of the star increases. This ignites the same nuclear reaction higher in the star in regions surrounding the core. At the same time, the diameter of the star increases enormously and its surface temperature cools, resulting in a *red giant* star. This phase in a star's life is relatively short, lasting about one tenth as long as the initial phase. As the newly formed helium falls to the core, the core continues to collapse and its temperature increases. For more massive stars, eventually the core becomes hot enough to fuse helium into carbon, oxygen, and, if there is sufficient mass, neon, and possibly heavier elements. During this period the star undergoes mass loss due to the low gravity in the higher altitudes of the star. This leads to the formation of a very short lived *planetary nebula* (about 10,000 years); see Figures 1 and 2 of the online supplement [van Dyk et al. (2009)].

In stars with initial mass less than about $8M_\odot$, the dense core eventually reaches a new equilibrium (a *degenerate electron gas*) that prevents further collapse even in the absence of a thermonuclear reaction. As the outer layers of the star blow away, eventually only a stable core composed of helium, carbon, and oxygen remains. These *white dwarf* stars are typically smaller than the Earth, are very dense (about one ton per cubic centimeter), and cool extremely slowly. Their lifetimes are measured in gigayears. For stars with an initial mass greater than $8M_\odot$, the degenerate electron gas does not prevent further collapse of the core. The continued collapse leads to higher and higher temperatures and the thermonuclear synthesis of progressively heavier elements. Eventually only degenerate neutron pressure stops the collapse, but not before the electrons of the atoms are forced into the atomic nuclei where they combine with protons to form neutrons and thus a *neutron star*. Matter falling into the newly formed neutron star sets off a shock wave that dramatically blows off the outer layers of the star in a *supernova* explosion; see Figure 3 and 4 of the online supplement [van Dyk et al. (2009)]. For even more massive stars, not even the degenerate neutron pressure can halt the collapse of the core. This leads to indefinite collapse and the formation of a *black hole*.

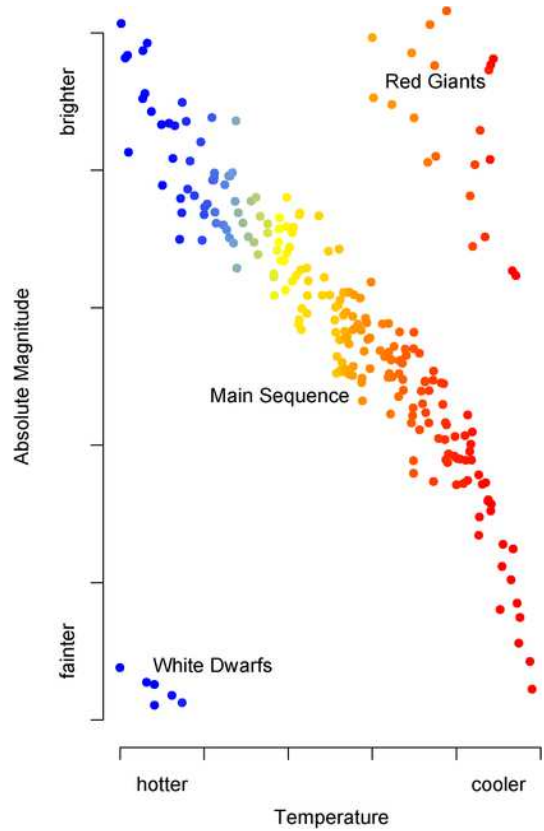

Fig. 1. *Schematic HR Diagram. The plot shows a schematic Hertzsprung–Russell (HR) diagram. The main sequence stars, red giants, and white dwarfs are all easily recognizable. The main sequence is broader than we would expect in a star cluster and more like we would expect to see with a star population that includes stars of different ages and metallicities.*

As a star evolves, its luminosity at different wavelengths of light changes. A Color-Magnitude Diagram (CMD) can be used to exploit this to identify stars at different stages of their lives. The original version of these diagrams, named for their inventors, are called Hertzsprung–Russell diagrams (HR diagram) and plot *absolute luminosity* on the vertical axis and stellar surface temperature on the horizontal axis. The absolute luminosity is the luminosity that the star would have if it were 10 parsecs (32.6 light years) away as opposed to the *apparent luminosity* it has when viewed from Earth. It is only possible to compute absolute luminosity of objects that are a known distance from Earth. A schematic example of an HR diagram appears in Figure 1. The stars labeled "Main Sequence" are stars in their initial phase of life when they have a hydrogen-burning core. There is a continuum of stars in this group that can be indexed by their initial masses. Stars to the upper left are more massive, hotter, and brighter.[2] These stars tend to burn their hydrogen more quickly, are shorter lived, and are the first to migrate to the group of stars labeled "Red Giants." Notice that the red giants are both cooler and more luminous. Their cooler temperatures make them appear redder while their massive sizes increase their luminosity. Finally, after a star loses its upper layers and its thermonuclear reaction fails, it migrates to the faint "White Dwarf" group at the bottom of the HR diagram.

HR diagrams are the oldest type of CMD, but there are many others. All CMDs are designed to use magnitudes in different color bands or *photometric magnitudes* to identify the evolutionary stages in the lives of stars. We generally simply refer to the set of photometric magnitudes as the magnitudes of a star. Because we focus on stellar clusters which consist of stars that are all nearly the same distance from Earth, we can use apparent luminosity in place of absolute luminosity and avoid the tedious task of determining the distance of each star. Second, the surface temperature of a star is highly correlated with the ratio of the star's luminosities in (nearly any) two optical color bands. (This corresponds to a difference in magnitudes, since magnitude is a logarithmic transformation of luminosity.) Thus, we need not directly determine the temperature of each star. Figure 2 illustrates the type of CMD we focus on. The data are from the Hyades cluster discussed in Section 5 and the plot compares the difference in apparent magnitudes (relative apparent luminosity) in the B band ("blue" containing violet, indigo, and blue light) and the V band ("visual" band containing cyan, green, and yellow light) on the horizontal axis with the apparent luminosity in the V band on the vertical axis. Just as in the HR diagram, the main sequence and white dwarfs are clearly visible. There are only a few giants at the top of the diagram. This is expected because stars spend a relatively short period of their lives as giants. Thus, the CMD has the same utility as the HR diagram in identifying the evolutionary groups, but without the absolute calibration.

---

[2]This relationship stems from the Stefan–Boltzman law for blackbodies (i.e., perfect radiators), which serves as a very good approximation for stellar radiation. The law says that absolute luminosity is proportional to radius squared times temperature to the fourth power.

We have seen that the initial mass of a star influences its location on the CMD. Initial composition is also important. Metallicity is a measure of the abundance of elements heaver than helium. These heavier elements tend to absorb light at the blue end of the spectrum and inhibit thermal (heat) radiation. Thus, stars with higher metallicity have a somewhat different set of colors and photometric magnitudes. Similarly, stars with more helium at their cores tend to have a less efficient thermonuclear reaction, simply because the hydrogen fuel is less pure. To compensate for this, the cores of these stars tend to be somewhat smaller, denser, and hotter. This in turn causes the stars to be more luminous and shorter lived, and again affects their colors and magnitudes. Two other variables affect the apparent magnitudes. A portion of the light from a star is absorbed by interstellar material. The more absorption and the farther a star is away, the less luminous it appears from Earth. Thus, six parameters, the initial mass, the metallicity, the helium abundance, the distance, the absorption, and the age of the star, determine a star's placement on the CMD. Exactly where it lands, however, requires complex physical calculations that are accomplished using sophisticated computer models.

2.2. *Computer-based stellar evolution models.* The computer-based stellar evolution models that we use to predict a star's placement on the CMD are a combination of several component computer models. In particular, there are a number of different computational implementations of computer models for main sequence and red giant stars. For this initial phase of stellar evolution, we use the state-of-the-art models by Girardi et al. (2000), by the Yale–Yonsei group [Yi et al. (2001)], and of the Dartmouth Stellar Evolution Database [Dotter et al. (2008)]. These models take the six parameters discussed in Section 2.1 as inputs and predict the placement of main sequence and red giant stars on the CMD. (Some of the models do not depend on helium abundance and thus have only five input parameters.) The main sequence/red giant models vary subtly in their implementation of the underlying physics and give somewhat different predictions. One of our primary goals is to compare these models empirically and to examine which, if any of them, adequately predict the observed data.

Unfortunately, all of these models break down in the turbulent last stage of red giants as they fuse progressively heavier elements at different shells of their interior, begin to pulsate, contracting and expanding, finally lose their outer layers in planetary nebulae and form white dwarfs. This transition is physically very complex and dominated by chaotic terms. Other computer models are used for white dwarfs. We use the white dwarf evolution models of Wood (1992) and the white dwarf atmosphere models of Bergeron, Wesemael and Beauchamp (1995) to convert the surface luminosity and temperature into magnitudes. Finally, to bridge the main sequence/red giant computer models with the white dwarf model, we use an empirical mapping that links the initial mass of the main sequence star with the mass of the resulting white dwarf [Weidemann (2000)]; this is the so-called *initial-final mass relation*. The combination of these several computer models for various stages of stellar evolution into one comprehensive stellar evolution model[3] was proposed by von Hippel et al. (2006).

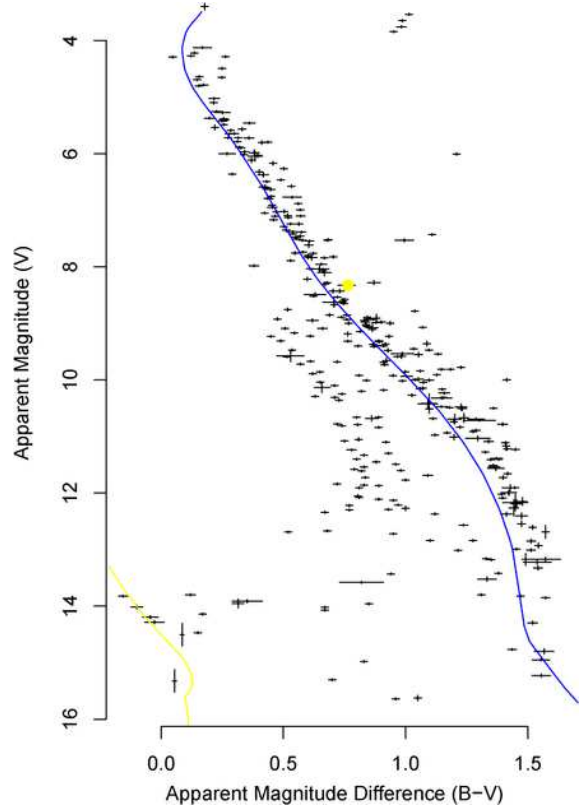

FIG. 2. *The Haydes CMD. The plot shows a Color-Magnitude Diagram (CMD) of stars in the Haydes cluster that we analyze in Section 5. Rather than artificially coloring the individual stars as in Figure 1, we plot all but one of them in black. The one yellow star is a binary star called vB022 that we discuss in Section 5. Each star is plotted with 95% intervals representing the measurement errors in $B-V$ and $V$. The star groups are less readily apparent than in Figure 1, largely because field stars contaminate the diagram. The swarm of stars below and to the left of the main sequence are field stars. These stars are mostly more distant and hence apparently fainter than the main-sequence stars. A small number of red giants appear in the upper center of the CMD. The units on the vertical axis are magnitudes, which are on a log scale with lower numbers indicating brighter sources. The units of the horizontal axis are differences in magnitudes. The blue and yellow lines are the fitted (Yale–Yonsei) main sequence and white dwarfs models.*

Thus, our stellar evolution model combines (i) the main sequence and red giant models with (ii) the initial-final mass relation, and (iii) white dwarf cooling models, to create a model we call the stellar evolution model, as it is meant to depict stars in all of the main phases of stellar evolution. (We ignore exotic objects such as neutron stars and black holes and short-lived objects such as supernovae, as the former objects do not radiate substantially in visible light and the latter objects are too short-lived to model sensibly from the color-magnitude diagram.) Here we avoid the details of the physics used in the computer-based stellar evolution model. Instead we refer interested readers to the basic description given in the online supplement to this paper [van Dyk et al. (2009)] and to the more detailed discussion that can be found in the many papers cited here and in the online supplement.

2.3. *Empirical exploration.* Our primary goal is to develop principled statistical methods that allow us to use observed data to fit the parameters of the stellar evolution models and to evaluate the empirical fit of these models. We focus on data that can easily be collected simultaneously on each star in a large field, in particular, on the intensity of each star's electromagnetic radiation in each of several wide wavelength bands, which we refer to as the star's magnitudes. We typically use two or three magnitudes for each star in our analysis, but could use ten or more. The number of stars in the data set can vary substantially, as few as 50 or as many as 50,000 are possible.[4] Currently we focus on data sets with fewer than 500 stars.

At least the brighter stars of most of the clusters we are studying are observable with small 1m-class telescopes. These instruments are common, are typically located in the desert southwest of the United States or in northern Chile, and can be equipped with cameras that focus approximately one square degree of the sky onto a charge coupled device (CCD) detector. The detector is sensitive to light with wavelengths of less than about one micron (infrared light) and either the detector or the atmosphere precludes sensitivity below about 350 nm (shortest wavelength of visible light). CCD detectors provide no information on the wavelength of this light, and so observations are made through filters that only allow wavelengths in a (typically) 100–200 nm band to be observed. Taking separate images through several filters allows us to observe several photometric magnitudes. For the faintest stars of interest, particularly the white dwarfs, we often need to employ the same techniques, but with 4–8 m class telescopes or with the Hubble Space Telescope.

---

[3]Our use of "stellar evolution model" is somewhat different than is in common use in the astronomical literature, where it generally refers to a model for the evolution of the main sequence and red giants. We use it to refer to a more comprehensive model that includes the transition to and evolution of white dwarfs.

[4]Open clusters are groups of up to a few thousand stars inside a galaxy that are loosely bound by gravity. Globular clusters, on the other hand, are tightly bound by gravity, composed of hundreds of thousands of stars, and are external satellites to a galaxy. Our current work focuses on open clusters which may have 50–500 stars cataloged in a data set.

Other types of observations are available to astronomers, but are typically more costly. For example, the metallicity of a star can be determined by careful analysis of a high-resolution spectrum, which is essentially thousands of photometric magnitudes recorded in very narrow wavelength bands for one star. The metallicity of a cluster can be determined by repeating this on ten or so stars in the cluster and comparing and combining the results. This requires much higher quality data than we are using. The results of previous analyses of this sort, however, can be used to formulate informative prior distributions for the metallicity parameter. Another example is the use of *proper motion* to determine which stars belong to a cluster. Proper motion is due to the relative motion of stars and our Solar System as they orbit the Galaxy and its measurement typically requires deep imaging that spans at least a decade. Only two dimensions of motion can be measured this way. Measuring an object's velocity along the line of sight (*radial velocity*) requires high-resolution spectral analysis. The electromagnetic waves from objects moving away from Earth are elongated, causing features in the visible spectrum to move toward the red end of the spectrum. The shift is known as the Doppler shift and can be measured for known spectral features and used to accurately measure radial velocity. A final example is the measure of distance using *parallax*. Objects that are relatively close to the Earth appear to make small movements on the sky as the Earth orbits the Sun. Precise knowledge of the diameter of the Earth's orbit along with simple geometry can be used to deduce the distance to the object. This method has been used to measure the distance to the stars in the Hyades cluster discussed in Section 5.

Although we typically observe several magnitudes for each star, the stellar evolution models are highly parameterized with five or six parameters for each star. Unfortunately, it is typically not possible to fit all of these parameters with useful accuracy using only a small number of magnitudes. To simplify the parameter space, we focus on stellar clusters. Not only do cluster stars possess nearly the same age, distance, metallicity, and helium abundance, but since the stars are moving through the galaxy as a group, their reddenings are also roughly the same. (Interstellar absorption is wavelength dependent and tends to absorb less red light, and hence reddens the appearance of the stars. The degree of reddening depends on the interstellar material, and hence the amount of absorption.) Thus, only the mass varies among the stars in a cluster and all other parameters are common to the

STATISTICAL ANALYSIS OF STELLAR EVOLUTION 11

cluster as a whole. As we shall see, this large reduction in the dimension of the parameter space makes it possible for us to satisfactorily fit stellar parameters.

Stellar surveys suggest that between one third and one half of all stars are actually binary or multi-star systems in orbit around their common center of mass. The stars in the majority of these systems are not directly distinguishable. For such systems, the observed luminosities are the sum of the luminosities of the component stars. (Magnitudes are on a log-luminosity scale and so must be transformed to luminosities before being added.) The added luminosity of the stellar companion tends to shift the star system up on the CMD, the larger the companion the greater the shift. If we do not properly account for this systematic distortion of the data, and, in particular, the location of the main sequence, it can bias the fitted stellar parameters. The fact that magnitudes of binaries are systematically different from nonbinaries and that the degree of this difference depends on the relative masses of the component stars, however, enables us to identify the component masses in a statistical model. Thus, we propose a model that accounts for unresolved multi-star systems. (Because what appears to be a star may actually be a multi-star system, we sometime use the words "star system" or simply "system." When there is no potential confusion, however, we continue to use the word "star" for these possibly multi-star systems.)

Field stars form a second type of contamination. As viewed from Earth, these stars are behind or in front of the stellar cluster, and thus are moving in a different direction as they orbit the Galaxy. Although careful measurements of proper motion and radial velocities can be used to determine if a star is moving with the cluster and thus help to distinguish cluster stars from field stars, such calculations have not been performed on all stars and are not always conclusive. Thus, we must build field star contamination into our statistical model.

## 3. A statistical model.

3.1. *Basic likelihood.* For a given set of stellar parameters, the stellar evolution model predicts a set of magnitudes. The observed magnitudes, however, are recorded with errors. Thus, we use the stellar evolution model to compute the mean structure used in a likelihood function and the distribution of the measurement error to model the variability. In particular, suppose we observe each of $N$ stars in each of $n$ filters. We denote the $N \times n$ matrix of observed magnitudes by $\mathbf{X}$, with typical element $x_{ij}$ representing the magnitude observed for star $i$ using filter $j$. We assume that the measurement errors follow a Gaussian distribution, $x_{ij} \sim N(\mu_{ij}, \sigma_{ij}^2)$, where $\mu_{ij}$ is the predicted magnitude under the stellar evolution model and $\sigma_{ij}^2$ is the variance of the measurement error, both for star $i$ using filter $j$. The means and variances also form $N \times n$ matrices, which we label $\boldsymbol{\mu}$ and $\boldsymbol{\Sigma}$.

TABLE 1
*Stellar parameters*

| Parameter | Astrophysical notation | Value |
|---|---|---|
| $\theta_{\text{age}}$ | $T$ | $\log_{10}$ age in $\log_{10}$ years |
| $\theta_{[\text{Fe/H}]}$ | [Fe/H] | Metallicity, $\log_{10}$ of the ratio of iron and hydrogen atoms[a] |
| $\theta_{[\text{He/H}]}$ | [He/H] | Helium abundance, $\log_{10}$ of the ratio of helium and hydrogen atoms |
| $\theta_{m-M_V}$ | $m - M_V$ | The difference between apparent and absolute magnitude[b] |
| $\theta_{A_V}$ | $A_V$ | Absorption in the V filter in magnitudes[c] |
| $M_{i1}$ | $M$ | Mass of the more massive star in binary-star system $i$ |
| $M_{i2}$ | $M$ | Mass of the less massive star in binary-star system $i$ |

[a]Iron is used as a proxy for all atoms heavier than helium because it is relatively easy to identify in spectral analysis. [Fe/H] is recentered using solar metallicity, so that a value of one means 10 time more iron relative to hydrogen than the Sun.

[b]The parameter $m - M_V$ is known as the *distance modulus*. Magnitude is a logarithmic measure of brightness, with smaller numbers corresponding to brighter objects. The difference between apparent and absolute magnitude depends on distance, which can be readily computed from the distance modulus. In particular, in the absence of absorption, $\theta_{m-M_V} = 5\log_{10}(d) - 5$, where $d$ is the distance measured in parsecs.

[c]The apparent magnitude in the V filter, $m_V$, can by computed from the absolute magnitude in the V filter, $M_V$, and $A_V$ via $m_V = M_V + A_V - 5\log_{10}(d) + 5$, where $d$ is the distance in parsecs.

While the components of $\boldsymbol{\Sigma}$ are assumed to be known from calibration of the data collection device, the components of $\boldsymbol{\mu}$ depend on the stellar parameters of interest via the stellar evolution model. Table 1 lists the model parameters. The first five rows in Table 1 list the stellar parameters that are common to all stars in the cluster and we refer to them as the *cluster parameters*. The only parameters that vary among the stars are their initial masses, $\mathbf{M}_1 = (M_{11}, \ldots, M_{N1})^\top$, where the subscript 1 indicates that we are assuming for the moment that the possibly multi-star systems are all unitary systems; see Section 3.2. If $\boldsymbol{\Theta} = (\theta_{\text{age}}, \theta_{[\text{Fe/H}]}, \theta_{[\text{He/H}]}, \theta_{m-M_V}, \theta_{A_V})$ is the vector of cluster parameters, then the expected magnitudes for star $i$ can be expressed as

$$\boldsymbol{\mu}_i = \mathbf{G}(M_{i1}, \boldsymbol{\Theta}), \tag{1}$$

where $\boldsymbol{\mu}_i$ is row $i$ of $\boldsymbol{\mu}$ and $\mathbf{G}$ is the $1 \times n$ vector-valued output from the stellar evolution model, with $G_j(M_{i1}, \boldsymbol{\Theta})$ the expected magnitude using filter $j$. For clarity, we refer to $\mathbf{G}$ as the stellar evolution model, and to the combination of the likelihood function and the prior distributions as the statistical model.

We can now write a preliminary likelihood function as

$$(2) \quad L_{\rm p}(\mathbf{M}_1, \boldsymbol{\Theta}|\mathbf{X}, \boldsymbol{\Sigma}) = \prod_{i=1}^{N}\left(\prod_{j=1}^{n}\left[\frac{1}{\sqrt{2\pi\sigma_{ij}^2}}\exp\left(-\frac{(x_{ij} - G_j(M_{i1}, \boldsymbol{\Theta}))^2}{2\sigma_{ij}^2}\right)\right]\right).$$

This likelihood was proposed by von Hippel et al. (2006) and we refer to it as the preliminary likelihood, using a 'p' in the subscript, because it does not account for binary-star systems or field star contamination, the subjects of the next two sections; see also DeGennaro et al. (2008). Although the Gaussian form of (2) is simple, the complex nonlinear function $\mathbf{G}$ cannot be expressed in closed form and complicates inference and computation.

One of our scientific goals is to compare and empirically evaluate individual and competing stellar evolution models. Thus, we may swap out $\mathbf{G}$ with a competing evolution model, say, $\boldsymbol{\mu}_i = \mathbf{H}(M_{i1}, \boldsymbol{\Theta})$ in (1) and (2).

3.2. *Binary stars.* For unresolved binary-star systems the observed luminosities are the sums of the luminosities from the two component stars. Because the relative masses of the component stars affect the observed magnitudes in a systematic way, it is possible to statistically identify both masses. Thus, we can construct a more sophisticated likelihood function that accounts for binary systems. In principle, the same is true of multi-star systems with more than two stars. Because these systems are significantly rarer than binary systems, and in the interest of parsimony, we confine our attention to binary systems.

We assume each star system has a primary and a secondary mass. The primary mass is the mass of the more massive component star and the secondary mass is zero if the system has only one star. Thus, let $\mathbf{M}$ be a $N \times 2$ matrix with typical row $\mathbf{M}_{i-} = (M_{i1}, M_{i2})$ representing the primary and secondary mass of star system $i$, respectively. Because the observed luminosities are simply the sum of the luminosities of the component stars, it is easy to modify the likelihood. Note, however, that the $L_{\rm p}$ is written in terms of magnitudes, which are on an inverted log-luminosity scale: magnitude $= -2.5\log_{10}(\text{luminosity})$. Thus, we simply replace (1) with

$$(3) \qquad \boldsymbol{\mu}_i = -2.5\log_{10}[10^{-\mathbf{G}(M_{i1}, \boldsymbol{\Theta})/2.5} + 10^{-\mathbf{G}(M_{i2}, \boldsymbol{\Theta})/2.5}]$$

and make a similar substitution in (2).

Because binary systems involving white dwarfs undergo a more complicated evolutionary history than we are able to model, we do not allow such binary systems in our model.

3.3. *Field star contamination.* Some of the stars in the observed field are not cluster members and, thus, their magnitudes are not well predicted by $\mathbf{G}$ evaluated at the cluster parameters. Each of these stars has its own value of $\Theta$ and we have no statistical power to identify all of these parameters. Thus, we assume a simple model for the magnitudes of the field stars that does not involve any parameters of scientific interest. In particular, we propose a uniform distribution on each of the magnitudes over a finite range that corresponds to the range of the data,

(4) $\quad p_{\text{field}}(\mathbf{X}_i) = c \quad \text{for } \min_j \leq x_{ij} \leq \max_j \text{ for } j = 1, \ldots, n,$

and zero elsewhere, where $\mathbf{X}_i$ is row $i$ of $\mathbf{X}$ and contains the observed magnitudes for star $i$, $(\min_j, \max_j)$ is the range of values for magnitude $j$, and $c = [\prod_{j=1}^n (\max_j - \min_j)]^{-1}$. Of course, a more sophisticated model could be used for the magnitudes of the field stars. For example, we could construct a nonparametric model using a wider field of stars, none of which are part of the cluster of interest. For our purposes, however, we find that this simple model does a good job of separating out stars that differ systematically from the cluster stars.

To construct the likelihood, we simply note that the observed data is a mixture of cluster stars and field stars, and use a two-component finite mixture distribution. In particular, we set $\mathbf{Z} = (Z_1, \ldots, Z_N)$, with $Z_i$ equal to one if star $i$ is a cluster member and zero if it is a field star. Thus, our final likelihood is

$$L(\mathbf{M}, \boldsymbol{\Theta}, \mathbf{Z} | \mathbf{X}, \boldsymbol{\Sigma})$$

(5)
$$= \prod_{i=1}^{N} \prod_{j=1}^{n} \left[ \frac{Z_i}{\sqrt{2\pi\sigma_{ij}^2}} \right.$$
$$\times \exp\left(-(\{x_{ij} + 2.5 \log_{10}[10^{-G_j(M_{i1}, \boldsymbol{\Theta})/2.5}\right.$$
$$\left. + 10^{-G_j(M_{i2}, \boldsymbol{\Theta})/2.5}]\}^2)(2\sigma_{ij}^2)^{-1}\right)$$
$$\left. + (1 - Z_i) p_{\text{field}}(\mathbf{X}_i) \right].$$

Our treatment of $\mathbf{Z}$ as a model parameter in the likelihood function is a departure from the standard practice of marginalizing (5) over $\mathbf{Z}$ in a finite mixture distribution. In a Bayesian analysis, however, it is natural to treat all unknown quantities in the same manner and, from a scientific point of view, we are sometimes interested in a particular star's cluster/field classification. Thus, we proceed with $\mathbf{Z}$ an argument of the likelihood function.

3.4. *Prior distributions.* We focus on a Bayesian analysis of this model at least in part because it allows us to directly incorporate prior information regarding the stellar parameters. We aim to accurately represent and quantify astronomical knowledge of likely values for the various parameters. For example, to reflect the fact that there are far more low mass stars than high mass stars, we use a Gaussian prior distribution on the base 10 logarithm

of the primary masses:

$$p(\log_{10}(M_{i1})) \propto \exp\left(-\frac{1}{2}\left(\frac{\log(M_{i1}) + 1.02}{0.677}\right)^2\right), \quad (6)$$

truncated to the range $0.1M_\odot$ and $8M_\odot$, where the constants are from the fit derived by Miller and Scalo (1979). (Note protostars with mass less than about $0.1M_\odot$ will not initiate a thermonuclear reaction and, for the clusters that we are interested in, stars with a mass greater than approximately $8M_\odot$ would have long ago evolved into a neutron star or a black hole, and thus would not be included in our data.) We use a uniform prior distribution on the unit interval for the mass ratio of the secondary mass over the primary mass. We need not truncate the secondary mass at $0.1M_\odot$ because low mass secondaries are taken as evidence for unitary systems.

Since the stars in a stellar cluster tend to move together, we can use proper motion, radial velocities, and, for nearby stars, parallax to help distinguish between cluster and field stars. For a well studied cluster such as the Hyades, these measurements are available for many stars and can be used to formulate prior probabilities for cluster membership; see Section 5. For less studied clusters, we may use a common prior probability based on the expected number of cluster stars. This can be estimated by simply comparing the number of stars per unit area in the cluster to areas nearby the cluster.

Turning to the cluster parameters, we use a uniform prior distribution between $\theta_{\text{age}} = 8.0$ and $\theta_{\text{age}} = 9.7$ for the $\log_{10}$ of age. This corresponds to a power law prior distribution on the age with exponent $-1$. We believe this distribution adequately reflects the observation that younger clusters are more common than older clusters. The remaining cluster parameters require cluster-specific prior distributions. We generally recommend putting Gaussian prior distributions on $\theta_{\text{[Fe/H]}}$, $\theta_{\text{[He/H]}}$, $\theta_{m-M_V}$, and $\log(\theta_{A_V})$. Informative prior distributions, however, require reasonable knowledge of the values and uncertainties of these parameters for a given cluster prior to analyzing the color-magnitude data. In our experience, informative prior distributions are not required for the cluster parameters [von Hippel et al. (2006); DeGennaro et al. (2008)]. Although in some cases narrow prior distributions help us better determine the likely values of the cluster parameters, we often find they are not needed for precise results.

### 4. Statistical computation.

4.1. *Basic MCMC strategy.* To fit the statistical model, we use an MCMC strategy. Each parameter is updated one-at-a-time in a Gibbs sampler. This is an ambitious strategy, given that there are $3N + 5$ free parameters in ($\mathbf{M}, \mathbf{\Theta}, \mathbf{Z}$) and strong linear and nonlinear correlations in the posterior distribution.

Owing to the complex form of the stellar evolution model, $\mathbf{G}$, none of the complete conditional distributions of $\mathbf{M}$ or $\mathbf{\Theta}$ are standard distributions or even available in closed form. Thus, each of these parameters is updated using a Metropolis rule with a uniform jumping rule, centered at the current value of the parameter being updated. Even this strategy is quite demanding because simply evaluating $\mathbf{G}$, and thus the target posterior distribution, is computationally very expensive.

To avoid evaluating $\mathbf{G}$ at every parameter update within each iteration, we use a tabulated version of $\mathbf{G}$ that is constructed before the MCMC run. The table has four dimensions corresponding to three of the dimensions of $\mathbf{\Theta}$ plus initial mass; absorption and distance modulus are handled differently. (Recall some models for main-sequence stellar evolution only require four cluster parameters. When we use these models the table is only three dimensional.) Each cell in the table records a vector of length $n$, corresponding to the expected magnitudes in each of the observed color bands. These are expected absolute magnitudes with no absorption, but can easily be converted to expected apparent magnitudes that account for absorption using the current values of $\theta_{m-M_V}$ and $\theta_{A_V}$. A typical table will include eight metallicity values, 50 ages, and about 190 initial masses. The values of these parameters are not evenly spaced and are chosen to capture the complexity of the underlying function. In fact, the number of mass entries may vary with age and metallicity depending on how complex the magnitudes are as a function of the initial mass. When evaluating the likelihood in the MCMC run, we use linear interpolation within the table to evaluate $\mathbf{G}$.

In one case we must extrapolate beyond the table. Unfortunately, the models for stellar evolution of the main sequence do not extend to masses less than $0.13 - 0.4 M_\odot$, depending on the stellar evolution model, metallicity, and age. This is an issue only for low mass companions, which are not the focus of our work. Moreover, for all but the smallest main sequence primary stars, a companion with mass less than $0.4 M_\odot$ makes little difference to the photometry of the system. Thus, we expect the relative accuracy of the extrapolation to be of little consequence for our overall fitted model. We do, however, want to allow for very small secondary stars, because many stars are in fact unitary. Thus, we extrapolate outside the tabulated model but do not trust the fitted masses or their error bars for small secondary stars, believing many of these systems to be simple unitary systems.

Overall, the use of a tabulated version of $\mathbf{G}$ substantially improves the computational performance of our sampler. One direct evaluation of $\mathbf{G}$, depending on the evolutionary state of the star, takes at least seconds on a modern desktop computer and could take more than an hour, while interpolating in 3 or 4 dimensions takes only a fraction of a second. With a table of high enough resolution, we gain a substantial amount computationally without significantly affecting the results.

In addition to simply evaluating the likelihood, there are a number of challenges in constructing the MCMC sampler so that its autocorrelations are not prohibitively high. For example, the posterior distributions of the stellar masses are highly dependent on whether a star is classified as a field star or a cluster star. In particular, the posterior distributions of the masses are much narrower for cluster stars, making it difficult for the sampler to change the field/cluster classifications when conditioning on the masses. We are able to reduce this correlation by using an alternative prior distribution on the masses of the field stars. Since these are nuisance parameters, this change has no substantive consequences. There are a number of other high linear and nonlinear posterior correlations among the continuous parameters. We eliminate these using a combination of static and dynamic transformations. These issues are discussed in the next three sections.

4.2. *Correlation reduction with an alternative prior specification.* As discussed in Section 3.3, because the values of the cluster parameters do not apply to the field stars, we are unable to constrain their masses using the stellar evolution model. Inspection of (5) reveals that $p(\mathbf{X}|\mathbf{M}, \boldsymbol{\Theta}, \boldsymbol{\Sigma}, \mathbf{Z})$ does not depend on the rows of $\mathbf{M}$ that correspond to stars classified as field stars. Put another way, if we condition on $Z_i = 0$ (i.e., star $i$ is a field star) for some subset of stars, the likelihood is not a function of the masses of those stars. Thus, for stars classified as field stars, the complete conditional distribution of their masses is simply the corresponding conditional distribution of the prior distribution. For stars classified as cluster stars, on the other hand, the likelihood can be very informative as to the masses and the complete conditional distribution of the masses may look very different. Simply stated,

$$p(M_{i1}, M_{i2}|\mathbf{X}, \boldsymbol{\Theta}, \boldsymbol{\Sigma}, \mathbf{Z})$$

is highly dependent on $Z_i$ and equal to $p(M_{i1}, M_{i2}|\boldsymbol{\Theta}, \boldsymbol{\Sigma}, \mathbf{Z})$ when $Z_i = 0$.

This dependence leads to intractable autocorrelations in the sampling chain. When the mass of a star that is classified as a field star is updated, it is unlikely to be valued in the range associated with cluster membership, even if the particular star has a substantial marginal posterior probability of cluster membership. Given enough iterations, the mass may migrate to the range associated with cluster membership, but the posterior relationship between $\mathbf{Z}$ and $\mathbf{M}$ nonetheless hampers efficient sampling.

To solve this problem, we take advantage of the fact that astronomers are only interested in the masses of stars that are cluster stars or, more precisely, in the conditional posterior distribution of mass given cluster membership. If we condition on cluster membership, that is, $Z_i = 1$ for each $i$, the choice of prior distribution for the masses of field stars is clearly irrelevant. Because the field star model does not depend on any of the parameters, we can further show that none of the posterior distributions of scientific interest are affected by the choice of $p(M_{i1}, M_{i2}|Z_i = 0)$. Namely, neither $p(\Theta, \mathbf{Z}|\mathbf{X}, \Sigma)$, $p(M_{i1}, M_{i2}, M_{j1}, M_{j2}, \Theta|Z_i = 1, Z_j = 1, \mathbf{X}, \Sigma)$, nor similar posterior distributions depend on the choice of prior distribution for the field star masses; see the appendix for details. Thus, how we sample the masses of field stars is immaterial to the final scientific analysis. From a sampling point of view, it would be ideal if the posterior distributions of the masses were identical regardless of the current cluster/field star classification. Since we are at liberty to set the conditional prior distributions of the mass given field star classification without upsetting our scientific conclusions, our strategy is to set this prior distribution so as to reduce the posterior relationship between $\mathbf{M}$ and $\mathbf{Z}$. The joint prior distribution can be factored via

$$p(\mathbf{M}, \Theta, \mathbf{Z}) = \prod_{i=1}^{N} p(M_{i1}, M_{i2}|Z_i) p(\mathbf{Z}) p(\Theta).$$

We continue to set the prior distributions $p(M_{i1}, M_{i2}|Z_i = 1)$ as in (6) and of $p(\mathbf{Z})$ and $p(\Theta)$ as described in Section 3.4. For $p(M_{i1}, M_{i2}|Z_j = 0)$, however, we use an estimate of $p(M_{i1}, M_{i2}|\mathbf{X}, Z_i = 1)$ based on its first two sample moments computed in an initial phase of the MCMC sampler. The estimate is parameterized as a $t_6$-distribution. Notice that this strategy does not mean that the complete conditional distributions of the components of $\mathbf{M}$ do not depend on $\mathbf{Z}$ because these distributions also condition on $\Theta$, but in our experience the dependence is weak enough to allow stars to efficiently switch from field to cluster and back.

A side effect of this prior specification affects the Metropolis acceptance probability when updating each of the $Z_i$. Because $p(M_{i1}, M_{i2}|Z_i)$ depends on $Z_i$, its values in the numerator and denominator of the acceptance probability will differ if the proposed value of $Z_i$ is different from the current value. This requires us to properly normalize this prior distribution, which can easily be accomplished analytically.

4.3. *Correlation reduction via static and dynamic transformations.* To avoid sampling inefficiency caused by high posterior correlations among the continuous parameters, we introduce a multivariate reparameterization. This involves both a simple static reparameterization of the masses and a dynamic reparameterization involving several parameters.

Since the total mass of the system is a principle determinant of the magnitudes, we expect the primary and secondary mass of each system to be negatively correlated. Preliminary analyses bore this out and suggested a static transformation that largely eliminates the nonlinear correlation. In particular, we define $R_i = M_{i2}/M_{i1}$ and use the ratio of the secondary mass to the primary mass in place of the secondary mass when constructing the sampler. We emphasize that this transformation removes nonlinear correlations: $M_{i1}$ and $R_i$ exhibit linear correlation in some cases. Our dynamic method for removing linear correlations is discussed below. When implementing the Metropolis update for each $R_i$, we reflect at the boundaries of the unit interval parameter space to maintain the symmetric jumping rule.

Preliminary analyses revealed a number of remaining strong linear correlations among the parameters. To adjust for these, we introduce a parameterized linear transformation of the parameter that is dynamically tuned to the strength of the correlation in a sequence of initial runs of the sampler. The functional form of the transformation is determined using a combination of astrophysics-based intuition and observation of the behavior of the sampler. Using a sequence of initial runs, we compute a mixture of conditional and marginal linear regressions on the sampled parameters. This sequence is generated in an ad hoc manner using trial and error to construct a transformation that is tuned to characteristics of the computer-based stellar evolution model and eliminates the large correlations in the Markov chain. The final transformation can be expressed as

$$M_{i1} = U_i + \beta_{R,i}(R_i - \widehat{R_i}) + \beta_{\text{age},i}(\theta_{\text{age}} - \widehat{\theta}_{\text{age}}) + \beta_{[\text{Fe/H}],i}(\theta_{[\text{Fe/H}]} - \widehat{\theta}_{[\text{Fe/H}]})$$
$$+ \beta_{m-M_V,i}(\theta_{m-M_V} - \widehat{\theta}_{m-M_V}),$$
$$\theta_{A_V} = V + \gamma_{[\text{Fe/H}]}(\theta_{[\text{Fe/H}]} - \widehat{\theta}_{[\text{Fe/H}]}) + \gamma_{m-M_V}(\theta_{m-M_V} - \widehat{\theta}_{m-M_V}),$$

where hats denote approximate posterior means that are calculated in an initial run for use in the transformation. The components of $\boldsymbol{\beta}$ and $\boldsymbol{\gamma}$ parameterize the transformation and are also computed during a sequence of initial runs using a sequence of simple linear marginal and conditional regressions; details are given in Section 4.4. The transformed variables, $U_i$ and $V$, are the residuals from these regressions. Thus, the MCMC sampler is run on the parameters $\{(U_1, R_1), \ldots, (U_N, R_N), \theta_{\text{age}}, \theta_{[\text{Fe/H}]}, \theta_{[\text{He/H}]}, \theta_{m-M_V}, V\}$, which we find significantly improves the convergence of the chain, as illustrated in the following section.

4.4. *Dynamic MCMC methods.* We begin the MCMC run with a burn-in period that is run with the transformed masses, but with the components of $\boldsymbol{\beta}$ and $\boldsymbol{\gamma}$ all set to zero. That is, the burn-in is run without the dynamic linear transformation. Upon completion of the burn-in, we implement a number of initial runs that are designed to compute components of $\boldsymbol{\beta}$ and $\boldsymbol{\gamma}$. After each of these runs, we update the definition of the $U_i$ or $V$ with the updated component of $\boldsymbol{\beta}$ or $\boldsymbol{\gamma}$. Thus, we begin with $\boldsymbol{\beta} = \mathbf{0}$ and $U_i^{(0)} = M_{i1}$ and regress $U_i^{(0)}$ on $(R_i - \hat{R}_i)$ to compute $\beta_{R,i}$ for each $i$. In this and all the regressions used to compute the dynamic transformations, the predictor variables are recentered at zero by subtracting off their means. Using the newly computed value of $\beta_{R,i}$, but with the other components of $\boldsymbol{\beta}$ still set to zero, we construct an updated transformation, $U_i^{(1)}$, that is used in place of $U_i^{(0)}$ in the second initial run. We continue in this way through the multiple initial stages that are described in Table 2. Notice that, in the first runs, we filter out stars that appear to be field stars and force the remaining stars to be classified as cluster stars. This results in an MCMC sampler that is more robust to poor starting values and poor choices of $\boldsymbol{\beta}$ and $\boldsymbol{\gamma}$, and can more easily find the posterior region of high mass. Once we have tuned the transformation parameters, we allow cluster-field star jumping of all stars in the data set and update all of the components of $\boldsymbol{\beta}$ and $\boldsymbol{\gamma}$. Some of the regressions are conditional and others are marginal. These choices were made via trial and error, with the aim of improving the mixing of the sampler. In some cases, when the fitted transformation parameters are small and not statistically significant and/or have a sign that is at odds with astrophysical intuition, we set the transformation parameter equal to zero. For example, $M_{i1}$ and $\theta_{\text{age}}$ are highly correlated for white dwarfs and largely unrelated for main-sequence stars. Thus, many of the $\beta_{\text{age},i}$ coefficients are fixed at zero.

The acceptance rates for the Metropolis jumping rules are monitored throughout the initial runs. If the acceptance rate among the previous 200 proposals falls below 20%, the width of the uniform jumping rule is decreased. If the rate grows above 30%, the width is increased. Initial run six in Table 2 is a period when only the acceptance rates are monitored.

The number of draws in the burn-in period and each of the initial draws can be set by the user. Currently we use 30 thousand draws in the burn-in and 5 or 10 thousand draws in each of the initial runs. Regression analyses are run using every fiftieth of the 5 or 10 thousand draws. This results in substantial computing time (typically half to three quarters of the total) being devoted to the burn-in and initial draws. As illustrated in Figure 3, however, this is a good investment. We are able to obtain nearly uncorrelated posterior samples and reliable summaries of a complex posterior distribution.

**5. The Hyades.** The Hyades is 151 light years away [Perryman et al. (1998)] and is the nearest star cluster to our Solar System.[5] The cluster

---

[5]The group of stars known as the Ursa Major Moving group is thought to be a dispersed cluster of stars formed from the same molecular cloud. The stars appear to have similar metallicity, age, and are moving as a group. At only 81 light years away and with its dispersed nature, this group of stars is scattered across a large portion of the northern sky and includes nearly all of the bright stars in the Big Dipper. The Sun is moving toward these stars, but at ten times the age is not part of this grouping.

Table 2
*Sequence of initial runs used to compute correlation reducing transformation*

---

In the initial burn-in period and in the first 6 initial runs each star's cluster membership status is held constant. That is, the $Z_i$'s are not updated from the starting values input by the user.

0. Burn-in period.
1. Compute each $\beta_{R,i}$ by regressing each $U_i^{(0)}$ on $R_i$.
2. Compute each $\beta_{\text{age},i}$ by regressing each $U_i^{(1)}$ on $\theta_{\text{age}}$. In this run $\theta_{[\text{Fe/H}]}$, $\theta_{[\text{He/H}]}$, $\theta_{m-M_V}$, and $\theta_{A_V}$ are fixed at our best estimate of their posterior means.
3. Compute each $\beta_{m-M_V,i}$ by regressing each $U_i^{(2)}$ on $\theta_{m-M_V}$. Compute $\gamma_{m-M_V}$ by regressing $V^{(0)}$ on $\theta_{m-M_V}$.
4. Compute each $\beta_{[\text{Fe/H}],i}$ by regressing each $U_i^{(3)}$ on $\theta_{[\text{Fe/H}]}$. Compute $\gamma_{[\text{Fe/H}]}$ by regressing $V^{(1)}$ on $\theta_{[\text{Fe/H}]}$.
5. Approximate the posterior mean and variance of $M_{i1}$ and $R_i$ to construct the alternative prior distributions on the masses for field stars.
6. Fine tune step sizes used in the Metropolis proposals to optimize acceptance rates.

In a second set of 7 initial runs, the above runs are repeated (including a second burn-in period), but this time the cluster memberships are sampled.

Step sizes for all parameters are adjusted continuously throughout all of the initial runs. Predictor variables are recentered at zero in all regressions.

---

is visible to the unaided eye and forms the nose of Taurus the Bull. The distance to the Hyades can be accurately computed using stellar parallax of its constituent stars. The age of the cluster has also been measured and is believed to be about $625 \pm 50$ million years [Perryman et al. (1998)]. This estimate is based on the fact that as a cluster ages its most massive stars are the first to evolve into red giants. These massive stars are at the upper left of the main sequence, the first part of the main sequence to disappear from a CMD. This so-called *main sequence turn off* can be used to estimate the age of a cluster [e.g., Chaboyer, Demarque and Sarajedini (1996); Montgomery, Marschall and Janes (1993); Sarajedini et al. (1999)]. Our primary scientific goal is to compare these age estimates with age estimates determined primarily from the colors and magnitudes of white dwarf stars. Since the cluster stars have a common age, we expect these age estimates to be similar. Up until now for the Hyades, however, the best age estimate based on white dwarfs [300 million years, Weidemann et al. (1992)] is about half the best estimate based on the main sequence turn off [625 million years; Perryman et al. (1998)]. Thus, the comparison is an opportunity to evaluate the underlying physical models and analysis techniques. To focus our analysis on white dwarfs, we remove both the red giants and the stars in the main sequence turn off from our data set.

More generally, we aim to evaluate our statistical method and the underlying computer models by comparing existing measurements with those obtained with our likelihood-based fit of the stellar evolution model and to compare the observed colors and magnitudes with those predicted by the stellar evolution models. This investigation is the most sophisticated empirical test of the computer-based stellar evolution models to date. Here we present only a sampling of our results. Detailed simulation studies under the simplified model given in (2) appear in von Hippel et al. (2006). More detailed comparisons of the stellar evolution models for the main sequence and discussion of the ramifications for the differences on the fitted stellar and cluster parameters appear in Jeffery et al. (2007) and DeGennaro et al. (2008).

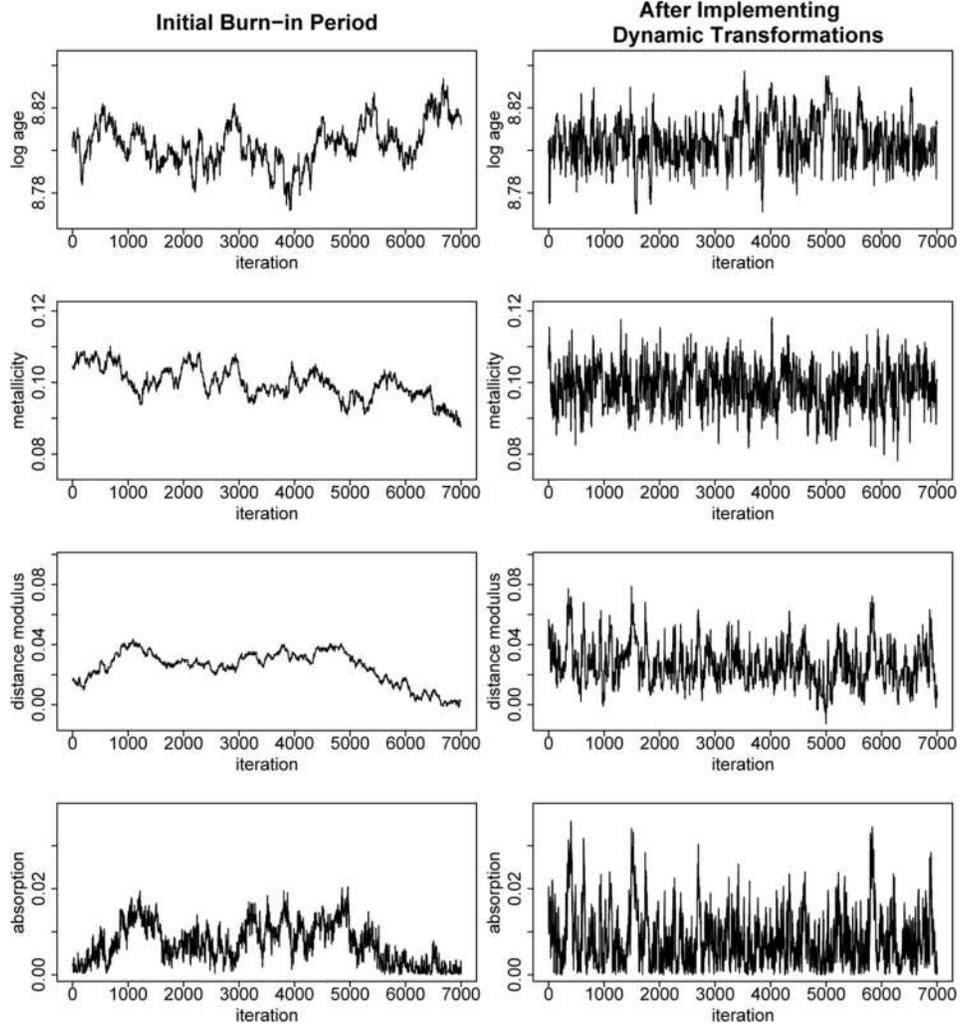

Fig. 3. *Improving convergence with Dynamic Transformations. The plots in the left column show time series plots of the MCMC draws of the four cluster parameters ($\theta_{\rm age}$, $\theta_{\rm [Fe/H]}$, $\theta_{m-M_V}$, and $\theta_{A_V}$, respectively) during the initial burn-in period. The plots represent a portion of the chain after it has reached the vicinity of the posterior mode but before the dynamic transformations are implemented. The right column shows time series plots of the same four parameters after the dynamic transformations have been computed and implemented. The transformations significantly reduce the autocorrelations of the chains.*

Figure 4 represents our fitted values for the $\log_{10}$ cluster age and cluster metallicity, $\theta_{\rm age}$ and $\theta_{\rm [Fe/H]}$. The two plots give 67% posterior intervals computed under the three stellar evolution models for the main sequence and compare them with the most reliable parameter estimates based on the main sequence turn off for age [Perryman et al. (1998)] and based on high-resolution spectral analysis for metallicity [Taylor and Joner (2005)]. Such best available estimates are used to formulate prior distributions for all cluster parameters except age. Because age is the parameter of primary scientific interest, we use a uniform prior distribution for $\theta_{\rm age}$; see DeGennaro et al. (2008) for an analysis of the sensitivity to the choice of prior distribution.

Because our goal is to estimate the age of the Hyades based on the colors and magnitudes of the white dwarf stars and because it is known that the stellar evolution models are flawed for the faintest main sequence stars (see the discrepancy between the observed magnitudes and the fitted Yale–Yonsei main-sequence model at the lower right of Figure 2), we repeat our analysis, leaving out stars with $V$ magnitudes fainter than a series of given thresholds. The horizontal axes of the two plots in Figure 4 are the magnitude of the faintest main sequence stars used in the analysis. It is apparent from the plots that as we include fainter stars in the analysis, the posterior distributions change considerably. It is also evident that the fitted values are quite sensitive to the choice of stellar evolution model for the main sequence stars. One of the primary aims of our study is to evaluate the reliability of the physics-based stellar evolution models. Figure 4 makes it clear that none of the models is reliable for the faintest stars.

Although our primary scientific goal is to determine $\theta_{\rm age}$ based on white dwarfs, some main sequence stars must be included to constrain the other cluster and stellar parameters. These parameters depend much more heavily on the main sequence data and models. Thus, in the left most fit in the lower panel of Figure 4, where only white dwarfs are included in the data set, the posterior and prior distributions for $\theta_{\rm [Fe/H]}$ coincide. As we include more data, the posterior distribution for $\theta_{\rm [Fe/H]}$ changes substantially and becomes more dependent on the choice of model. The cluster age, however, is far less sensitive to the choice of model, at least for stars of magnitude about 8.5 and brighter. Thus, despite the inaccuracies and/or approximations in the stellar evolution models for the main sequence, we are able to reliably estimate the age and for the first time produce a white-dwarf age estimate that agrees with the most reliable age estimate based on the main sequence turn off.

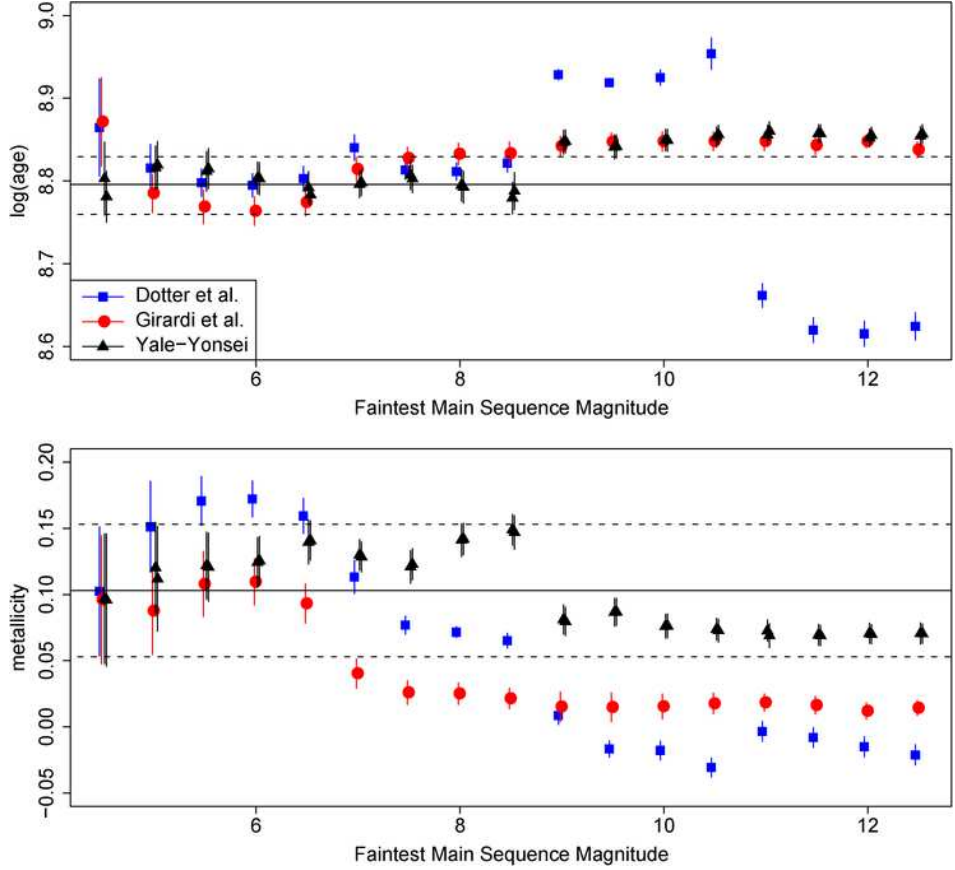

Fig. 4. *Effect of Data Depth on Fitted Age and Metallicity. The two plots show the posterior mean and one posterior standard deviation intervals for $\theta_{\rm age}$ and $\theta_{\rm [Fe/H]}$, respectively. The horizontal axis indicates the faintest magnitude of main sequence stars included in the data; recall that higher magnitudes correspond to fainter stars. The fit is repeated using each of the three stellar evolution models for main sequence stars. The model compiled in the Dartmouth Stellar Evolution Database [Dotter et al. (2008)] is represented by blue squares, the model of Girardi et al. (2000) by red circles, and the Yale–Yonsei model by black triangles. The Yale–Yonsei model is replicated with two sets of starting values. The black horizontal lines are the mean and one standard deviation intervals for the most reliable external estimate of the age and metallicity of the Hyades. This estimate was used to quantify the Gaussian prior distribution on $\theta_{\rm [Fe/H]}$, while a flat prior distribution was used on $\theta_{\rm age}$. The plots show that the stellar evolution models break down for the faintest stars and under-represent uncertainty in the fits.*

The sensitivity of the fitted values both to the choice of stellar evolution model and to the depth of data included in the analysis clearly indicate that the posterior standard errors computed under any particular model are underestimates of the actual uncertainty for the cluster parameters. This is true for $\theta_{\rm age}$ as well as the other parameters. Systematic errors stemming from apparent inaccuracies and/or approximations in the stellar evolution models contribute substantially to the uncertainty. A synthesis of the information in Figure 4 into a best estimate of cluster parameter along with a reliable estimate of uncertainty is of particular scientific interest to an astronomer. A formal statistical approach might use model averaging to combine the perspectives of the three stellar evolution models into a single coherent analysis. One might expect the resulting posterior variance to be larger under the mixed model than under any of the individual models. A statistical analysis, however, is only as good as the model it is predicated upon. Thus, a better long-run strategy is to explore the differences among the stellar evolution models in light of the observed data, with the goal of designing models that more reliably represent the underlying physical processes and are better able to predict the observed data. For the time being, we base our final parameter estimates on main sequence stars of magnitude 8.5 or brighter and conduct a simple ANOVA-type analysis that combines the within-model and between-model uncertainty.

As a second evaluation of the underlying physical models, we compare the posterior distribution of the primary and secondary masses of a known binary star system called *vB022* to an externally computed estimate. The posterior distribution of the stellar masses computed using the Yale–Yonsei main-sequence evolution model and using main sequence stars down to magnitude 8.5 appears in Figure 5. The non-Gaussian character of the distribution is both striking and typical of many of the low dimensional marginal distributions. This highlights an advantage of our Bayesian approach: We are able to marginalize to the parameters of direct scientific interest in a natural manner that avoids any Gaussian approximation to the likelihood function.

To evaluate the underlying physical models, we compare the posterior distribution in Figure 5 to an external estimate of the stellar masses computed using the radial velocities of the two component stars in the system [Peterson and Solensky (1988)]. These measurements are quite reliable and are independent of the data and models that go into our estimates. Although our estimate of the secondary mass is consistent with the external estimate, the more reliable external estimate of the primary mass is about 5% larger than our estimate. We attribute this to systematic errors in the underlying physical models that we use. In Figure 2, *vB022* is marked by a yellow point, and its $V$ magnitude is 8.5. This is right at the point where the stellar evolution models begin to diverge in their fit; see Figure 4. This divergence grows worse lower in the main sequence; see Figures 2 and 4.

**6. Discussion.** We have described a Bayesian model-based approach to fitting the stellar and cluster parameters of physics-based computer models for stellar evolution. Our method constitutes the first statistical attempt to empirically evaluate and compare these models. Although initial results point to some inadequacies in the underlying models, their predictions do largely agree with the observed data. Thus, our white-dwarf based estimates of the cluster parameters are the best estimates available to date of their kind. That these estimates largely agree with the main sequence turn off estimates validates both our estimates and our technique, and, to a certain extent, the underlying computer models. In the future we hope that our technique can be improved with an extension of our methods to include red giants and main sequence stars at the turn off and by incorporating updated computer models. The larger and more informative data sets should yield even more precise estimates. Moreover, with more reliable computer models in hand, more sophisticated techniques, such as Bayes factors and model averaging, can be used to evaluate and compare the underlying physical models.

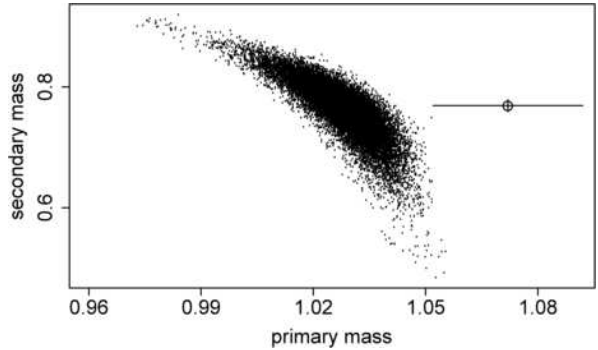

Fig. 5. *The Joint Posterior Distribution for the Primary and Secondary Masses of the Star vB022 in the Hyades. The scatter plot shows the Monte Carlo sample from the posterior distribution under our analysis using the Yale–Yonsei model for main sequence evolution. The star has a posterior probability of cluster membership equal to 99.955% and the plot gives the conditional posterior distribution of the two masses given that the binary system is a member of the cluster. This is compared with an external estimate of the two masses that is indicated by the open circle with whiskers that correspond to 95% marginal confidence intervals. Our estimate of the primary mass is about 5% lower than the more reliable external estimate. This difference is attributed to systematic errors in the underlying physical models. The masses are in units of $M_\odot$.*

## APPENDIX

In this appendix we verify that the posterior distribution of scientific interest is not affected by the choice of prior distribution for the stellar masses conditional on a star being a field star, namely, $p(\mathbf{M}_{i-}|Z_i = 0)$ for $i = 1, \ldots, N$, with $\mathbf{M}_{i-}$ the pair of masses for star $i$. In the interest of brevity, we rewrite the likelihood given in (5) as

$$(7) \quad L(\mathbf{M}, \mathbf{\Theta}, \mathbf{Z} | \mathbf{X}, \mathbf{\Sigma}) = \prod_{i=1}^{N} [Z_i f_1(\mathbf{X}_i, \mathbf{M}_{i-}, \mathbf{\Theta}) + (1 - Z_i) f_0(\mathbf{X}_i)],$$

where $\mathbf{X}_i$ is the vector of magnitudes observed for star $i$, $f_1$ is the joint distribution of the magnitudes for a cluster star, and $f_0$ is the joint distribution of the magnitudes for a field star. The joint posterior distribution can then be written

$$(8) \quad p(\mathbf{M}, \mathbf{\Theta}, \mathbf{Z} | \mathbf{X}, \mathbf{\Sigma}) \propto \prod_{i=1}^{N} [Z_i f_1(\mathbf{X}_i, \mathbf{M}_{i-}, \mathbf{\Theta}) p(\mathbf{M}_{i-} | Z_i) p(Z_i)$$
$$+ (1 - Z_i) f_0(\mathbf{X}_i) p(\mathbf{M}_{i-} | Z_i) p(Z_i)] p(\mathbf{\Theta}).$$

Expanding the product results in $2^N$ terms of the form

$$(9) \quad p(\mathbf{\Theta}) \prod_{i \in \mathcal{I}_1} Z_i f_1(\mathbf{X}_i, \mathbf{M}_{i-}, \mathbf{\Theta}) p(\mathbf{M}_{i-} | Z_i) p(Z_i)$$
$$\times \prod_{i \in \mathcal{I}_0} (1 - Z_i) f_0(\mathbf{X}_i) p(\mathbf{M}_{i-} | Z_i) p(Z_i),$$

where $\mathcal{I}_0$ and $\mathcal{I}_1$ partition $\{1, 2, \ldots, N\}$. (The $2^N$ terms correspond to the $2^N$ two-set partitions of $\{1, 2, \ldots, N\}$.) Due to the leading factor in each product, $p(\mathbf{M}_{i-} | Z_i) p(Z_i)$ is evaluated at $Z_i = 1$ for $i \in \mathcal{I}_1$ and at $Z_i = 0$ for $i \in \mathcal{I}_0$.

To compute the marginal posterior distribution $p(\mathbf{\Theta}, \mathbf{Z} | \mathbf{X}, \mathbf{\Sigma})$, we integrate (8) over $\mathbf{M}$, which corresponds to the sum of $2^N$ integrals over terms of the form given in (9). Because $\int (1 - Z_i) f_0(\mathbf{X}_i) p(\mathbf{M}_{i-} | Z_i) p(Z_i) d\mathbf{M}_{i-} = (1 - Z_i) f_0(\mathbf{X}_i) p(Z_i)$ for any proper choice of $p(\mathbf{M}_{i-} | Z_i)$, however, each of these integrals depends on $p(\mathbf{M}_{i-} | Z_i)$ only if $i \in \mathcal{I}_1$. Thus, $p(\mathbf{\Theta}, \mathbf{Z} | \mathbf{X}, \mathbf{\Sigma})$ does not depend on the choice of $p(\mathbf{M}_{i-} | Z_i = 0)$.

Using a similar argument, we can show, for example, that $p(\mathbf{\Theta}, \mathbf{M}_{i-}, \mathbf{M}_{j-} | Z_i = Z_j = 1, \mathbf{X}, \mathbf{\Sigma})$ does not depend on the choice of $p(\mathbf{M}_{i-} | Z_i = 0)$ for $i = 1, \ldots, N$. If we integrate (8) over the masses of a subset of the stars, the resulting distribution does not depend on $p(\mathbf{M}_{i-} | Z_i = 0)$ for the marginalized stars by the same argument as outlined above. For the remaining stars, $p(\mathbf{M}_{i-} | Z_i = 0)$ again falls out when we condition on their cluster membership, for example, $Z_i = Z_j = 1$.

**Acknowledgment.** We thank Elizabeth Jeffery for many helpful conversations.

## SUPPLEMENTARY MATERIAL

**Statistical analysis of stellar evolution: online supplement** (DOI: [10.1214/08-AOAS219SUPP](10.1214/08-AOAS219SUPP); .pdf). This supplement contains four color figures and a description of the physics behind the computer-based stellar evolution models. This material was originally intended to be included in this article, but was removed for editorial reasons. The images are visually impressive but not central to our statistical analysis. The section on the computer model provides details for readers interested in the inner workings of the likelihood function used in this article.

D. A. van Dyk  
2206 Bren Hall  
Department of Statistics  
University of California, Irvine  
Irvine, California 92697-1250  
USA  
URL: [www.ics.uci.edu/˜dvd](www.ics.uci.edu/˜dvd)  
E-mail: [dvd@ics.uci.edu](dvd@ics.uci.edu)

S. DeGennaro  
Department of Astronomy  
University of Texas at Austin  
1 University Station C1400  
Austin, Texas 78712-0259  
USA  
E-mail: [deg@astro.as.utexas.edu](deg@astro.as.utexas.edu)

N. Stein  
Department of Statistics  
Harvard University  
Cambridge, Massachusetts 02138-2901  
USA  
E-mail: [nmstein@fas.harvard.edu](nmstein@fas.harvard.edu)

W. H. Jefferys  
Department of Mathematics and Statistics  
University of Vermont  
Burlington, VT 05401  
USA  
E-mail: [bill@bayesrules.net](bill@bayesrules.net)

T. von Hippel
Physics Department
Siena College
515 Loudon Road
Loudonville, New York 12211
USA
E-mail: tvonhippel@siena.edu